\documentstyle[11pt]{article}
\makeatletter

\gdef\@publabel{\hfil}
\gdef\@pubdate{\null}
\gdef\@pubnumber{\null}
\gdef\@author{\null}
\gdef\@title{\null}
\gdef\@abstract{\null}
\long\def\pubdate#1{\gdef\@pubdate{#1}}
\long\def\pubnumber#1{\gdef\@pubnumber{#1}}
\long\def\publabel#1{\gdef\@publabel{#1}}
\long\def\author#1{\gdef\@author{#1}}
\long\def\title#1{\gdef\@title{#1}}
\long\def\abstract#1{\gdef\@abstract{#1}}
\def\titlerelax{
\let\maketitle\relax
\let\settitleparameters\relax
\let\consolidatetitle\relax
\let\inittitlepage\relax
\let\finishtitlepage\relax
\let\titlepagecontents\relax
\let\multithanks\relax
\let\titlebaselines\relax
\let\@makepub\relax
\let\@maketitle\relax
\let\@makeauthor\relax
\let\@makeabstract\relax
\let\@maketitlenote\relax
\let\thanks\relax
\let\titlerelax\relax}
\def\titleclean
{\gdef\@titlenote{}
\gdef\@abstract{}
\gdef\@author{}
\gdef\@title{}
\gdef\@pubdate{}\gdef\@pubnumber{}\gdef\@publabel{}
\gdef\@dpublabel{}
}

\def\@makepub{\vbox to \z@{\hbox to \textwidth{\hfill
\@publabel \hfill
\llap{\parbox[t]{0.33\textwidth}{\raggedleft\@pubnumber}}}%
\vss}}
\def\@maketitle{\vskip 60pt \begin{center}
 {\LARGE \@title \par}
 \end{center}}

\def\@makeauthor{{%
\def\and{\smallskip {\normalsize \rm and\smallskip }}
\def\And{\medskip {\normalsize \rm and\\}\medskip}
\long\def\address##1{{\def\and{\\and\\}\medskip
				{\small  \\##1\\}
}}
{\centering
 \vskip 3em
 \large \lineskip .75em
 \@author}
 \par}}
\def\@makedate{\vskip 1.5em
 {\raggedright \small \noindent\@pubdate \par}}
\def\@makeabstract{\vskip 1.5em
{\small
\begin{center}
{\bf ABSTRACT\vspace{-.5em}\vspace{0pt}}
\end{center}
\quotation \@abstract \endquotation}}
\def\maketitle{\titlepage
\let\footnotesize\small \setcounter{page}{0}
\@makepub
\vfil
\@maketitle
\@makeauthor
\vfil
\@makeabstract
\@thanks
\vfil
\@makedate
\if@restonecol\twocolumn \else \eject \fi
\titlerelax \titleclean
\setcounter{footnote}{0}
}

\pubnumber{UCLA/94/TEP/10 \\ hep-ph/9403313}
\pubdate{March 1994}
\title{Hadronic Spectrum in Inclusive Semileptonic B Decays}

\author{Antonio O.\ Bouzas\protect\thanks{Internet: bouzas@physics.ucla.edu} 
and  Dario Zappal\`{a}\protect\thanks{Internet: zappala@physics.ucla.edu}
\address{UCLA Department of Physics\\
Los Angeles, California 90024-1547}
}

\abstract{We evaluate the inclusive semileptonic B decay spectrum as a function
of the final-state hadrons energy to second order in the inverse quark mass 
expansion and tree level in
$\alpha_{s}$. 
We argue that there is an energy
interval below the $c$ production threshold that could be used to
determine $V_{ub}$. 
}

\newcommand{\bea}{\begin{eqnarray}}
\newcommand{\eea}{\end{eqnarray}}
\newcommand{\be}{\begin{equation}}
\newcommand{\ee}{\end{equation}}

\begin{document}

\maketitle

{\em 1.Introduction.\/} The inclusive semileptonic decay of the B
meson is of great importance for the determination of the Cabibbo-%
Kobayash-Maskawa (CKM) matrix elements $V_{cb}$ and $V_{ub}$. For
this reason several attempts have been made to predict the total width
and the electron energy distribution in this decay.  It has recently
been shown \cite{chay} that both can be computed in the framework of the QCD
operator-product expansion, without resorting to models for the
structure of the initial hadron.  To lowest order the parton model
result is recovered.  Matrix elements of higher dimension operators
are then expanded in powers of the inverse of the $b$ quark mass $1/m$
by applying heavy-quark effective theory (HQET) \cite{effm}.

This method has been applied \cite{bigi,mawi} (see also \cite{chay,mann})
to compute the  electron energy
spectrum to order $1/m^{2}$. However, for values of the electron
energy in the vicinity of its maximum kinematical limit the
corrections to the parton model are unreliable, being dominated by
large unphysical oscillations and furthermore singular at the
endpoint.  This fact spoils the possibility of determining $V_{ub}$,
since this quantity must be extracted from the endpoint region where
the $b\rightarrow c$ channel is kinematically forbidden. Some
tentative solutions to this problem have been suggested such as
gaussian smearing \cite{mawi} or integration \cite{bigi} of the spectrum in the
endpoint region. Also a form factor formalism has been proposed
\cite{neub}.

In this letter we compute the inclusive semileptonic differential
decay width as a function of the total energy of the final-state
hadrons employing the approach outlined above. We suggest that this
quantity could be suitable for the determination of $V_{ub}$. In this
case, the relevant part of the spectrum corresponds to total hadronic
energies smaller than the charm mass where $b\to u$ is the only decay
mode available and it is worthwhile to study the behaviour of the
heavy-quark approximation scheme within this region.

\hspace{0.5cm}

{\em 2. Hadronic spectrum.\/} In order to obtain the hadronic energy
distribution we closely follow the procedure used in \cite{mawi} for the
electron spectrum.  For that reason, we omit all technical details
related to the computation.  In the sequel we neglect radiative
corrections of order $\alpha_{s}/\pi$.

We start from the weak hamiltonian
\be\label{hamil}
{\cal H}(x)=-V_{jb}\frac{G_{F}}{\sqrt 2} J^{\mu}_{l}(x) J_{\mu} (x)
\ee
where $G_{F}$ is the Fermi constant, $V_{jb}$ is the CKM matrix element for 
$b \rightarrow j$
and $J_{l}^{\mu}$ and $J_{\mu}$ are the leptonic and hadronic weak currents 
respectively,
\bea
J^{\mu}_{l}(x)  & = &  \overline{\psi_{e}}(x)\gamma^{\mu} (1-\gamma_{5})\psi%
_{\nu}(x)\label{cur1}\\
J_{\mu}(x)    & = &  \overline{\psi_{j}}(x)\gamma_{\mu} (1-\gamma_{5})\psi_{b}
(x)\label{cur2}
\eea
The squared amplitude for the semileptonic B decay can be written as
($k$ and $k'$ are the electron and neutrino momenta respectively, 
and $q=k+k'$),
\be \label{ampl}
|A|^2={{|V_{jb}|^2 G_F^2}\over{2}} (2\pi) L^{\mu\nu}(k,k')W_{\mu\nu}(q) 
\ee
The leptonic tensor $L_{\mu\nu}$ appearing in (\ref{ampl}) is given at 
tree level by the ex\-pres\-sion%
\footnote{We use the convention $\epsilon_{0123}=1$.},
\bea
\lefteqn{ \sum_{pol} \langle l,\overline\nu| J^{\nu}_l(x) |0\rangle   
\langle 0| J^{\dagger\mu}_l(0) |l,\overline\nu\rangle   =  e^{iqx} 
L^{\mu\nu}(k,k')}\nonumber\\
 &  & \hspace{2 cm} = e^{iqx}\,\, 8 \left( k^\mu {k'}^\nu + {k'}^\mu  k^\nu 
- (k \cdot 
k') g^{\mu\nu}
  -i\epsilon^{\mu\nu\alpha\beta} k_{\alpha} {k'}_{\beta} \right)\label{lete}
\eea
The hadronic tensor, defined as,
\be\label{hate}
W_{\mu\nu}(q)= \frac{1}{2 \pi} \sum_X \int d^4 x e^{iqx}
\langle B|J^{\dagger}_{\mu} (0) |X\rangle   \langle X| J_{\nu}(x) |B\rangle
\ee
(where the sum runs over a complete set of hadronic states) can be
related to the discontinuity of the matrix element of the time ordered
product of hadronic weak currents,
\be\label{time}
T_{\mu\nu}=-i \int d^4 x e^{iqx} \langle B| T J^{\dagger}_{\mu}(0) J_{\nu}(x) 
|B\rangle
\ee
through the relation,
\be\label{rela}
\mbox{Im} \left( T_{\mu\nu}L^{\mu\nu}\right) =-\pi W_{\mu\nu}L^{\mu\nu}
\ee
Finally, using  (\ref{ampl}) and (\ref{rela}), the B decay width is given by 
\be\label{widt}
\Gamma_B=|V_{jb}|^{2} G_F^{2}\int {{d^4 k}\over{(2\pi)^4}}
{{d^4 k'}\over{(2\pi)^4}}
\delta(k^2)\theta(k_0)\delta({k'}^2)\theta({k'}_0)
\mbox{Im}\left( - T_{\mu\nu}L^{\mu\nu}\right)
\ee
with the normalization
$\langle B(M_{B}v)| B(M_{B}v^{\prime})\rangle = (2\pi)^{3} v^{0} 
\delta^{3}(M_{B}v -M_{B}v^{\prime})$  \cite{mawi}, $M_{B}$ and $v$ 
being the mass
and velocity respectively of the B meson.
The problem of calculating the total decay width and the leptonic and
hadronic spectrum  is then related to the evaluation of the tensor 
$T^{\mu\nu}$. We shall limit ourselves to the case $j=u$ and the
$u$ quark mass will be neglected. 

Due to the $T$-ordering the matrix element in (\ref{time}) contains a 
$u$ quark propagator.
In the following we shall treat separately two contributions
to the quantity $T_{\mu\nu}L^{\mu\nu}$
\be\label{contr}
T_{\mu\nu}L^{\mu\nu}=T^{(1)}_{\mu\nu}L^{\mu\nu}+
T^{(2)}_{\mu\nu}L^{\mu\nu}
\ee
where the first term contains the lowest order approximation for the $u$ 
propagator ({\em i.\frenchspacing e.} the free propagator) whereas in the 
second one
the first order correction to the propagator, consisting of a gluon insertion,
is taken into account.

The first step to compute $T^{(1)}_{\mu\nu}L_{\mu\nu}$ is to 
factorize out the large $x$-dependence in the b field,
\be\label{fase}
\psi_b(x)=e^{-imvx}\psi_b'(x)
\ee
with $m$ being the $b$ quark mass and $v$ the 4-velocity of the B meson.
The field on the right-hand side of (\ref{fase}) can be split into 
two components $\psi_b'(x)=h(x)+\chi(x)$ defined by the properties
$\not\hspace*{-0.5 ex}v h(x)=h(x)$ and $\not\hspace*{-0.5 ex}v \chi(x)=
- \chi(x)$.
$\chi(x)$ can be eliminated in terms of $h(x)$ through an expansion in 
powers of $1/m$ \cite{effm}.

At the same time, we can perform a short-distance expansion since the 
$x$ dependence of $\psi_b'(x)$ is small in the sense
that it is related to the ``residual'' momentum $\kappa$ defined as,
\be\label{resi}
\kappa=p_b-mv
\ee
$\kappa$ is non zero for finite $m$ because the $b$ quark does not carry 
all of the
B momentum, but it vanishes in the limit of very large quark mass.

The combination of short-distance expansion and HQET leads to a 
$1/m$ expansion of the matrix 
element in terms of operators of increasing dimensions.
The lowest order coincides with the 
the semileptonic decay of a free $b$ quark, 
since it corresponds to setting $\kappa=0$, ({\em i.\frenchspacing e.},
$p_b=mv$) and thus identifying the decaying particle with the
$b$ quark. Corrections of order $1/m$ vanish \cite{chay},
whereas those of order $1/m^2$ are expressed in terms of the two 
matrix elements \cite{bigi,mawi}
\bea
K_b & = & -\langle B|\overline h(0) {{(iD)^2}\over{2 m^2}} h(0)|B\rangle
                                                         \label{condk}\\
G_b & = & Z_b\langle B|\overline h(0) {{gG^{\alpha\beta}\sigma_{\alpha\beta}}
\over{4 m^2}} h(0)|B\rangle\label{condg}
\eea
$D$ being the QCD covariant derivative, $g$ the strong coupling constant,
$\sigma_{\alpha\beta}=(i/2)[\gamma_\alpha, \gamma_\beta]$, 
and $Z_bG^{\alpha\beta}$ the renormalized gluon field strength tensor.
The quantity $G_b$ is directly related to the experimentally measured
mass splitting $M(B^{*})-M(B)$ from which one gets 
the small value $G_b=-0.0065$.  The numerical value of $K_{b}$ is estimated 
to be
$K_b=0.01$ \cite{bigi,mawi} from  QCD sum rules (see \cite{bigi}).

The explicit expression for $T^{(1)}_{\mu\nu}L^{\mu\nu}$ to the order
$1/m^2$ is given by,
\be\label{expl}
T^{(1)}_{\mu\nu}L^{\mu\nu}=64\int d^4 x \int{{d^4 p}\over{(2\pi)^4}}
e^{-i(mv-q-p)x} {1\over{p^2+i\varepsilon}}
(R_0 + R_1 +R_2 )
\ee
where  $p$ is the $u$-quark momentum and,
\bea
R_{0} & = &  (p\cdot k) ~ (v\cdot k')\nonumber\\
R_{1} & = &  i ~ m (K_b+G_b) ~  x_\rho \left({2\over3}{k'}^\rho -
{5\over 3}(v\cdot k')~v^\rho\right)(p \cdot k)\label{erre}\\
R_{2} & = & {1\over 3}m^2 K_b  (p\cdot k) ~ (v\cdot k')  
\left(x^2-(v \cdot x)^2 \right)\nonumber
\eea
The discontinuity of $T^{(1)}_{\mu\nu}L^{\mu\nu}$ in (\ref{expl}) is 
given by the 
$u$ quark propagator
\[ 
\mbox{Im} \left({1\over {p^2+i \varepsilon}}\right) =-\pi \delta(p^2)
\]
The integration over $x$ in (\ref{expl}) yields terms proportional to 
$\delta^{4}(mv-q-p)$
and its first two derivatives,
according to the power of $x$ appearing in $R_i$.

We notice that each term $R_i$ of the expansion is symmetric in the two 
momenta $p$ and $k$. This symmetry can be related to a Fiertz identity
for the product of currents in the amplitude that exchanges the electron and
quark fields $\psi_{e}$, $\psi_{u}$.
Thus, from (\ref{widt}), (\ref{expl}) and (\ref{erre}), 
the first piece of the decay width $\Gamma^{(1)}_{B}$ is invariant
under exchange of $p$ and $k$ and
the two spectra ${{d\Gamma_B^{(1)}}/{d E_e}}$ and
${{d\Gamma_B^{(1)}}/{d E_u}}$  have exactly the same form.

After performing the integrations we obtain,
\bea
{1\over{\Gamma}}{{d\Gamma_B^{(1)}}\over{d w}} & = & 
\left[ 6w^2-4 w^3 - {{20} \over 3} K_b w^3  - 
{{20} \over 3} G_b w^3 +4 G_b w^2 \right] \theta (1-w)\nonumber\\
&  & +
\left[ {2\over 3} K_b +{{10}\over{3}} G_b \right] \delta(1-w)+
{2\over 3} K_b \delta '(1-w)\label{wid2}
\eea
where we introduced 
the dimensionless variable $w=2E_{u}/m$ and the lowest 
order width $\Gamma=(|V_{ub}|^{2}G_{F}^{2}m^{5})/(192\pi^{3})$.

Let us now consider the second term $T_{\mu\nu}^{(2)}L_{\mu\nu}$ in 
(\ref{contr}). 
The complete expression for $T_{\mu\nu}L^{\mu\nu}$ 
to order ${1/{m^2}}$ includes the correction obtained by attaching
a gluon line to the free $u$ propagator. 
The insertion of the gluon field on the propagator in the matrix element 
of (\ref{time}) generates a term proportional to $G_b$ \cite{mawi}.  
Notice that this insertion is not related to the
one loop self-energy correction to the propagator, whose imaginary part is
proportional to the probability of real gluon emission.
The new term is, then,
\bea
\lefteqn{T^{(2)}_{\mu\nu}L^{\mu\nu}=\frac{128}{3}\int d^4 x 
\int{{d^4 p}\over{(2\pi)^4}}
e^{-i(mv-q-p)x} \Bigl ({1\over{p^2+i\varepsilon}} \Bigr )^2
m^2 G_b\times}  \nonumber \\
&  & \hspace{5 cm}\left[(v\cdot k)~ (p\cdot k') - (k\cdot k')~ (v\cdot p)
\right] \label{newt}
\eea
Unlike $T_{\mu\nu}^{(1)}L_{\mu\nu}$, in the above equation the expression in 
square 
brackets
is antisymmetric in $p$ and $k$.

In fact the presence of the gluon field in the hadronic current breaks
the symmetry between $\psi_e$ and $\psi_u$. In this case one can still
apply a Fiertz transformation exchanging $\psi_e$ and the string
$\psi_u (\overline \psi_u G \psi_u)$. This operation yields again the result
(\ref{newt}), but it does not simply correspond to the exchange 
of $p$ and $k$ in the 
scalar 
products of (\ref{newt}). It also has the effect of 
substituting $-G_{b}$ for $G_{b}$.

In (\ref{newt}) the discontinuity generated by the squared propagator is,
\[
\mbox{Im} \Bigl ({1\over {p^2+i \varepsilon}}\Bigr )^2 =\pi \delta'(p^2)
\]
Due to the integration over $p_0$, the derivative acting on the delta 
function can be transferred to the other terms in (\ref{newt}). This way
$\mbox{Im} (T^{(2)}_{\mu\nu}L^{\mu\nu})$ contains no derivatives of
$\delta(p^2)$ but the derivative of the momentum conservation delta function
has been generated.
Thus, we can see that both in the first part of the calculation 
in (\ref{expl})  and in the gluon emission part (\ref{newt}), the potentially 
singular terms appear at this step of the computation as
derivatives of $\delta^{4}(mv-q-p)$.

The contribution of $T^{(2)}_{\mu\nu}L^{\mu\nu}$ to the spectrum, obtained 
by integrating in (\ref{newt}) is, 
\be\label{wid3}
{1\over{\Gamma}}{{d\Gamma_B^{(2)}}\over{d w}}=
{4\over 3} G_b \Bigl [ (4 w  - 3 w^2) \theta(1-w) - \delta (1-w) \Bigr ] 
\ee
As already mentioned, the gluon contribution to the leptonic spectrum 
has a different form. From (\ref{newt}) one gets
($y=2 E_e/ m$),
\be\label{wid4}
{1\over{\Gamma}}{{d\Gamma_B^{(2)}}\over{d y}}=
{4\over 3} G_b \Bigl [ - 9 y^2 ~\theta(1-y) + 3 \delta (1-y) \Bigr ] 
\ee
Incidentally, we notice that both (\ref{wid3}) and (\ref{wid4}), although
relevant in the decay energy spectra, give zero contribution 
to the total B decay width.

Finally, the $u$ quark decay spectrum to the order $1/m^2$ is given 
by the sum of (\ref{wid2}) and (\ref{wid3}) and it is plotted in Fig.\  1. 
Due to 
the presence of the delta functions we have limited the range of $w$ 
to $0<w<0.9$ in the figure, and to show the small $K_b$
dependence 
curves for $K_b=0$, and $0.05$ (dotted curves) are also included.
At $w=0.5$ we find that the differential width changes by about 5\%
as $K_{b}$ varies from 0 to 0.05.

\hspace{0.5cm}

{\em 3.Discussion.\/} The differential decay rate for the process
$B\to e\overline \nu X$ is obtained, as shown above, from the
discontinuity of $T_{\mu\nu}L_{\mu\nu}$ whose cut is given by the
propagator of the final $u$ quark (or $c$ quark if $X$ contains
charmed mesons) only.  It is natural, then, to identify the total
energy of the final-state hadrons $E_{X}$ with the energy carried by
the $u$ ($c$) which is the only particle contributing to the imaginary
part of the amplitude.  This would obviously not be true had we
included $O(\alpha_{s}/\pi)$ radiative corrections.  In this case the
propagator would be dressed by self-energy corrections, so that for
instance a real gluon emission would give a new contribution to the
discontinuity.

Let us now briefly comment on our results.
Equations (\ref{wid2}) and (\ref{wid3}) show that the hadronic 
spectrum has the same kind
of singularities at the endpoint $w=1$ that characterize the electron
spectum at $y=1$.  To explain this point one should notice that both
hadron and electron spectra for the decay of a free $b$ quark 
(that is our lowest order approximation) do not vanish smoothly
at the endpoint $w=1$ and $y=1$ but contain functions $\theta(1-w)$
and $\theta(1-y)$.  Singularities appear in the following corrections
due to expansion of those step functions \cite{bigi,neub}. More explicitly,
looking at (\ref{expl}) in momentum space, we just have the expansion of
$\delta^4(p_b-p-q)$ around $\kappa=0$
\be\label{delt}
\delta^4(p_b-p-q)=\delta^4(mv-p-q) + \kappa\cdot\partial 
{\delta^4}(mv-p-q) +... 
\ee
and performing the integration, due to the symmetry of the variables
$p$ and $k$, the derivatives in (\ref{delt}) are transformed in derivatives of
$\delta(1-w)$ when computing ${{d\Gamma_B}/{dw}}$ or of $\delta(1-y)$
for ${{d\Gamma_B}/{dy}}$, thus generating analogous singularities for
the two cases. The same argument is valid for the singularity in the
gluon insertion correction (\ref{newt}) since, as already discussed, it can be
expressed again as a derivative on $\delta^4(mv-p-q)$.

Another point to be discussed concerns the region around $w=0$.
As explained in \cite{chay} the discontinuity of $T^{\mu\nu}L_{\mu\nu}$ 
relevant to the 
problem considered here is given, in the ($v \cdot q$) complex plane
at fixed $q^2$, by a cut along the real axis in the range 
$\sqrt {q^2}<v\cdot q < (M^2_B+q^2-M_h^2)/(2 M_B)$
($M_h$ is the mass of the lightest hadron that can be produced
in the $B$ decay). For large values of $q^2$ this cut gets
closer to another cut of the amplitude, corresponding to different physical 
processes, defined for $v\cdot q>((2 M_B+M_h)^2 -q^2 -M_B^2)/(2M_B)$. 
More precisely when $q^2$ takes its maximum value $(M_{B}-M_h)^2$, the 
first cut is 
reduced to one point $v\cdot q =(M_{B}-M_h)$ and the other starts at 
$v\cdot q=M_B+3M_h$. In our case $M_h$ is the pion mass and
we must expect the breakdown of the expansion when $q^2$
is maximum. We remark that this problem is not related to the $\delta$ and
$\delta^{\prime}$ 
in (\ref{wid2}) and (\ref{wid3}), those appear as soon as we allow 
$\kappa\neq 0$ independently
of the value of $q^2$.
In our calculation no hadronic  mass $M_{h}$ appears and
since all the decay products are massless, the maximum value of
$q^2$ is $m^2$. In the electron spectrum we have $q^2=m^2$ at the 
endpoint $E_e=E_e^{max}$.  Conversely, in the case of
${{d\Gamma_B}/{dw}}$, we find $q^2=m^2$ at the opposite 
edge of the spectrum that is at $w=0$.

{}From (\ref{wid2}) and (\ref{wid3}), since $G_b$ is negative, we see that
the spectrum takes negative values around $w=0$ in a region of the 
same order of magnitude as $G_b$. Although this is a very small effect
because of the value of $G_b$, it is clear that our spectrum 
is a poor approximation to the physical one which, as a matter 
of fact, must be zero for any value of the energy below the first threshold.

With no information about the subsequent terms in the series, one can
reasonably take $\Lambda_{QCD}$ as an estimate of the size of the
energy range where the expansion breaks down, and eventually consider
the spectrum in (\ref{wid2}), (\ref{wid3}) as a realistic 
approximation for energies
above, say, $.8$ Gev. Therefore, we suggest the possibility of using
the hadronic spectrum to determine $V_{ub}$. In fact $b\to c$
contributes to the B decay only for energies greater than the charm
mass, so that $b\to u$ can be safely considered the only allowed
channel up to approximately 1.1 GeV.  This means that there is an
energy interval left, of about the same size as the one available at
the endpoint of the electron spectrum, where the hadronic energy
distribution can be of interest, being free from the problem of
singularities that affects the electron spectrum at $E_e=E_e^{max}$.

In conclusion we have evaluated the B decay spectrum as a function 
of the massless final quark energy at the $1/m^2$ level and to lowest
order in $\alpha_s/\pi$ and we have found that it shows, at the maximum
energy endpoint, the same kind of singularities as the electron spectrum 
and, at the same time, it turns out to be unphysical also in the low energy 
region that is typically dominated by non perturbative effects. Still
it seems that there is a small range, usable for the $V_{ub}$ 
determination, where the approximation holds.

\hspace{0.5cm}

{\em Acknowledgements.\/} We would like to thank R.\ Peccei
for suggesting this problem. 
We benefited from many helpful discussions with G.\ Baillie, B.\ Hill and 
K.\ Wang.

A.\ O.\ B.\ is supported by an ICSC World Laboratory scholarship.
D.\ Z.\  is supported by INFN.

\newpage

\section*{Figure Caption.}

Differential width $d\Gamma/dw$ as a function of $w$ as given by 
the sum of (\ref{wid2}) and (\ref{wid3}). The value of the
parameters is $G_{b}=-0.0065$ and $K_{b}=0.01$ (solid line), $K_{b}=
0.0$ (upper dotted curve), $K_{b}=0.05$ (lower dotted curve).

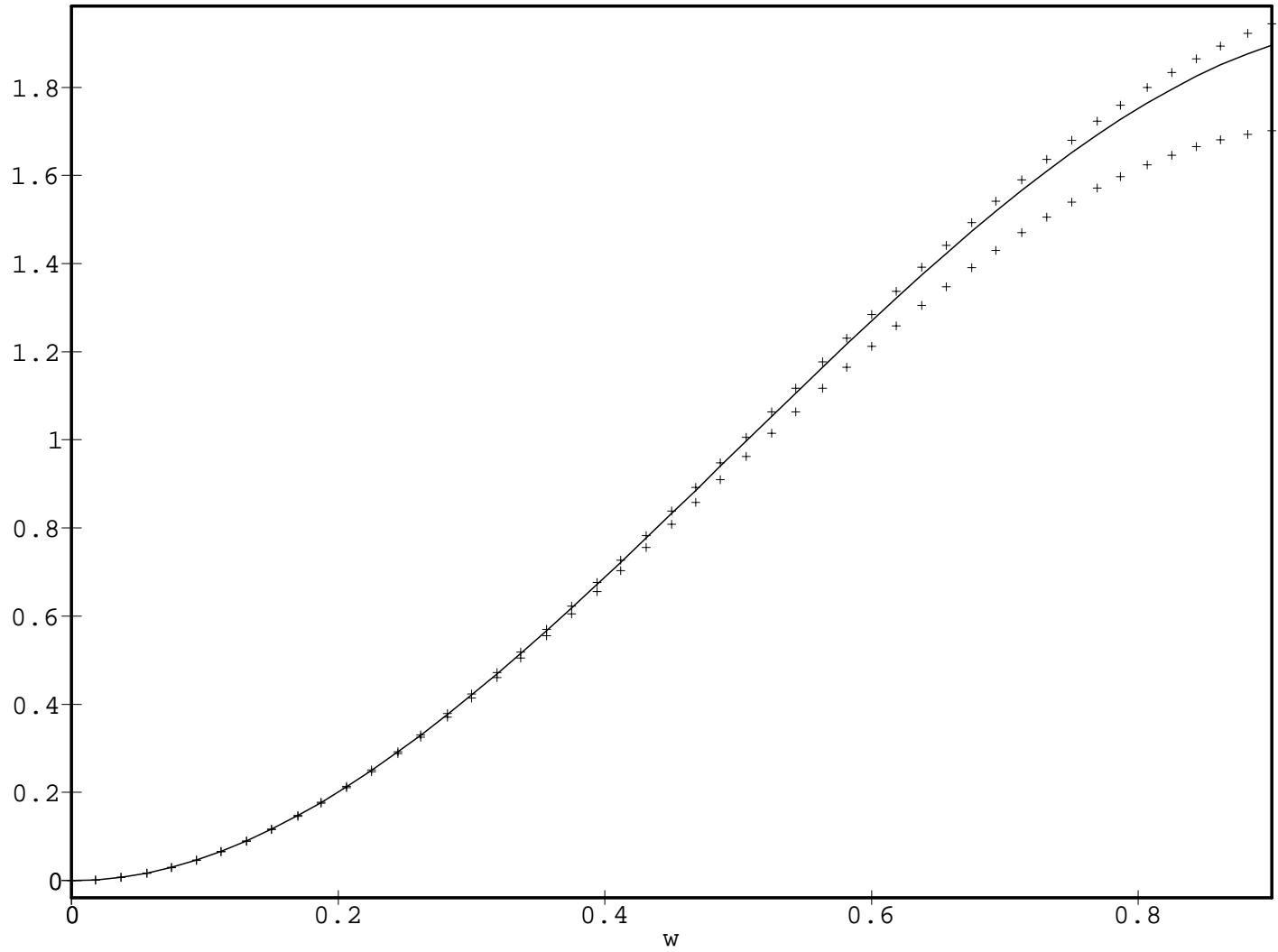
\end{document}